\documentclass[prd, twocolumn, nofootinbib, floatfix]{revtex4-2} 

\usepackage{amsmath}
\usepackage{graphicx}
\usepackage{dcolumn}
\usepackage{bm}
\usepackage{epsfig}
\usepackage{amssymb,latexsym,mathrsfs,empheq}
\usepackage{graphicx}
\usepackage{color}
\usepackage{xcolor} 
\usepackage{mathtools} 
\usepackage{hyperref} 
\usepackage{cleveref} 
\usepackage{soul}

\hypersetup{
    colorlinks=true,
    linkcolor=red,
    citecolor=blue,
} 

\newcommand{\be}{\begin{equation}}
\newcommand{\ee}{\end{equation}}
\newcommand{\bs}{\begin{split}} 
\newcommand{\bea}{\begin{eqnarray}}
\newcommand{\eea}{\end{eqnarray}}

\newcommand{\om}{\Omega_m} 
\newcommand{\ode}{\Omega_{\rm de}}
 
\newcommand{\wde}{w_{\rm de}} 
\newcommand{\wtot}{w_{\rm tot}} 

\newcommand{\weff}{w_{\rm eff}}

\newcommand{\rde}{\rho_{\rm de}}

\begin{document}

\title{Null Impact of the Null Energy Condition in Current Cosmology} 

\author{Robert R.\ Caldwell${}^{1}$ \& Eric V.\ Linder${}^{2}$} 
\affiliation{
${}^1$ Department of Physics and Astronomy,
Dartmouth College, Hanover, NH 03755, USA\\ 
${}^2$ Berkeley Center for Cosmological Physics \& Berkeley Lab, 
University of California, Berkeley, CA 94720, USA
} 

\begin{abstract} 

We clarify the role of the oft-misunderstood Null Energy Condition (NEC) in the context of the current cosmological data. 
In particular, the NEC implies the sum of the total energy density and pressure satisfies $\rho_{tot}+P_{tot} \ge 0$; the energy conditions do not apply separately to individual components of the cosmological fluid. 
Consequently, we show that under the current best-fit cosmological model no violation of the NEC takes place, past or future. 
Further, growth in the energy density of an individual component cannot be used to signal violation of the NEC. 
We illustrate these points with a worked example whereby misestimation of the matter density leads to a phase during which $\rho_{de} + P_{de} < 0$ for the effective dark energy, followed by a phantom crossing and subsequent $\rho_{de} + P_{de} > 0$. At no time is the NEC violated. 
We also introduce ``elephant'' and ``chimera'' classes of physics for crossing $w_{de}=-1$. 
\end{abstract} 

\date{\today} 

\maketitle

\section{Introduction} 

Current cosmological data point toward a beautifully bizarre situation. Not only has the expansion of the Universe begun to {\it accelerate}, but the {\it accelerant} itself has peaked \cite{desidr2,desidr2lya,desidr2de}. 
In terms of the ratio of pressure to energy density, $w \equiv P/\rho$, the equation of state of the total cosmological fluid has crossed from $w_{tot} > -1/3$ to $w_{tot} < -1/3$, marking the onset of acceleration. The equation of state of dark energy 
(i.e.\ 
everything contributing to expansion other than matter
and radiation)
has crossed the so-called phantom divide, from $w_{de} < -1$ to $w_{de} > -1$, implying that dark energy density was growing but is now decaying. 

As if the acceleration was not preposterous enough, the behavior of the accelerant is even more baffling due to the paucity of physical mechanisms that can produce phantom dark energy, $w_{de} < -1$. Even fewer physically motivated systems can manifest both an episode of $w_{de}>-1$ and $w_{de} < -1$. And if phantom dark energy were to dominate the cosmological fluid
and thereby violating 
the Null Energy Condition (NEC), there is danger of a future, Big Rip singularity. 

To clarify several of these points we discuss various issues related to $w_{de}<-1$, $\wtot<-1$, evolution of $\rho_{\rm de}$, and the NEC, showing that nothing necessarily pathological is implied by the current best fit cosmology. Section~\ref{sec:nec} discusses some aspects of the NEC, while Section~\ref{sec:cross} examines the mechanisms for and implications of $w_{de}<-1$ and $w_{de}$ crossing $-1$. A simple example of crossing by confusion among dark sectors is worked through in Section~\ref{sec:dm}. We apply these lessons to the best fit current cosmological data in Section~\ref{sec:data} and conclude in Section~\ref{sec:concl}.

\section{Null Energy Condition} \label{sec:nec} 

In order to appreciate the role of the NEC in cosmology, it is important to understand how the NEC is defined and how it is used. 
The energy conditions (e.g.\ \cite{2003.01815,Curiel:2014zba,1401.4024}) -- weak, strong, dominant, and null (to name a few) -- are a set of inequalities that reasonable forms of matter are thought to satisfy. Their primary application is in theorems about the formation of singularities in gravitational collapse and in cosmology \cite{Penrose:1964wq,Hawking:1970zqf}. Under the various energy conditions, we can make general statements about the convergence of geodesics, existence of trapped surfaces, the attractiveness of gravity, and singularities.

The NEC requires $R_{\mu\nu}k^\mu k^\nu \ge 0$ for any null 4-vector $k$. Using Einstein's Equations to relate the Ricci curvature to the stress-energy tensor expressed as an ideal fluid, we find $\rho_{tot} + P_{tot} \ge 0$. Hence, the NEC is a 
condition on the sum of all forms of stress-energy. 
For homogeneous, isotropic cosmology, this means the entire fluid. Similarly, the Strong Energy Condition (SEC) requires $R_{\mu\nu}t^\mu t^\nu \ge 0$ for any time-like 4-vector $t$. In terms of an ideal fluid, $\rho_{tot} + 3 P_{tot} \ge 0$ must be satisfied by the total fluid. 

Although these conditions may be a good idea in general, they fail in certain relevant cases and are by no means sacrosanct. The SEC fails for a homogeneous, isotropic spacetime undergoing accelerated expansion -- whether at present or in the early Universe. Furthermore, all of the above energy conditions tend to fail in the presence of quantum fields. A standard example is the stress-energy due to the Casimir effect, which violates the NEC and Dominant Energy 
Conditions \cite{qph0106045,hth0703067,1709.01999,2301.02455}. 
As an aside, consideration of the role of quantum fields in gravitation has prompted the introduction of Quantum Energy Inequalities and Averaged Energy Conditions that aim to restrict the duration or spatial extent of various energy condition violations \cite{2003.01815,gr-qc/0001099,2503.19955}.

From the perspective of a cosmologist, the energy conditions only refer to the background, homogeneous energy density and pressure, and say nothing about the behavior of fluctuations. If the fluctuations behave reasonably, then phantom-like dark energy  (or dark energy that crosses $w_{de}=-1$) may be feasible. It is important to distinguish between a scenario that violates the NEC, which is a statement about the total stress-energy tensor, and a component that is phantom-like, with $w_{de} < -1$.

The quantity $\rho+P$ also appears in the continuity equation
\be 
\dot\rho=-3H(\rho+P) \ , 
\ee 
whereby $\rho+P<0$ implies that $\dot\rho>0$, and vice versa. Hence, violation of the NEC implies growth of the total energy density in an expanding spacetime. Consequently, the increase of the dark energy density over time 
is sometimes interpreted as a violation of NEC. However, this is not true since $\rho\ne\rho_{de}$, i.e.\ there are other components entering the determination of the total stress-energy tensor and the structure of spacetime. 

A coupling between two species $i,\,j$ that transfers energy $i \to j$ can be arranged to cause decrease in $i$ and increase in $j$. This happens all the time, both cosmologically and terrestrially. 
But the increase over time of some component cannot be interpreted as NEC violation; it is the whole system that must be in violation.

\section{Phantom and Phantom Crossing} \label{sec:cross} 

Two aspects of dark energy should be distinguished: whether 
$w_{de}<-1$ and whether $w_{de}$ crosses $-1$, from above or below. A canonical single scalar field in 
general relativity has neither: it is restricted to $w \ge-1$ 
and is called quintessence. 
(For simplicity, $w$ without a subscript will refer to $w_{de}$.) 
To enable $w<-1$ requires either 
changing the kinetic structure away from canonical or going 
beyond general relativity. Multiple quintessence fields cannot 
give $w<-1$, so if one has multiple fields then at least one 
of them must violate one of the two conditions. 

Changing the kinetic structure gives, 
for example, the class of k-essence 
fields, and these can have $w<-1$. 
However, $w<-1$ can bring 
with it pathologies such as ghosts (runaway behavior often 
associated with a Hamiltonian unbounded from below, i.e.\ 
negative energy states). So one needs to check this carefully 
(see \cite{0811.0827} for a review). 
In any case, k-essence fields alone cannot cross $w=-1$ 
\cite{vikman05}. 

One solution to crossing $w=-1$ is having multiple fields, 
one k-essence and one quintessence, say. Indeed this can  
stabilize a ghost arising from a k-essence field alone \cite{2108.06294}. 
The multiple fields can be treated at the background (expansion) 
level as a single field with 
\be 
w_{\rm eff}=w_1\,\frac{\delta H^2_1}{\delta H_1^2+\delta H_2^2}+w_2\,\frac{\delta H^2_2}{\delta H_1^2+\delta H_2^2}
\, \label{eq:wsum} 
\ee 
i.e.\ the effective dark energy density weighted sum of the individual equations 
of state. 
(Here $\delta H_i^2$ is the contribution of component $i$ to the Friedmann equation, i.e.\ Hubble expansion rate squared or effective energy density.) 
This effective single field approach was developed 
in \cite{0410680,0507482,0808.3125,1008.1684} and is frequently 
used in Boltzmann codes to calculate perturbative quantities 
as well. For perturbations, the scalar sound speed plays a 
critical role. Interestingly, only certain behaviors for $w$ 
approaching $-1$ and the sound speed allow stable crossing of 
$w=-1$ (see Appendix B of \cite{1305.6982} for details). 

While this approach of using both quintessence and k-essence 
fields can enable crossing of $w=-1$, in the simplest approach 
it can only cross from above, i.e.\ from $w>-1$ in the past to $w<-1$ in the 
future (the opposite of what the data requires as we will see 
in Section~\ref{sec:data}). This is easy to see: if the phantom 
field dominates at early times then since $w_{\rm phantom}<-1$ 
implies that $\dot\rho_{\rm phantom}>0$ then the phantom field 
(if it always stays phantom) 
will always dominate and no crossing will occur. If the  
quintessence field dominates at early times, then the crossing 
goes from $w>-1$ in the past to $w<-1$ in the future. 

Thus, a physical mechanism to cross $w=-1$ in the way the data indicates suggests the need for multiple, interacting quintessence and k-essence fields. 
This ``multiplication 
of entities'' raises questions of Occam's razor. We can think 
of two classes of such elaborated theories, what we call elephants 
and chimeras. 


The well known legend of the elephant is that a group of 
scientists are led into a dark room and each describes what 
they detect. One finds something like a snake, one a fan, one 
a wall, one a tree trunk, one a rope. But these multiple elements 
are part of a unified framework - an elephant (trunk, ear, side, 
leg, tail). A unified framework can lead to further tests that 
establish the framework. 

Some examples of the elephant class include unified models 
with extra degrees of freedom such as spintessence (with extra 
scalar degrees of freedom) \cite{0105318,hep-th/0501160}, generalized 
Proca theories (with extra vector degrees of freedom; but 
crossing $w=-1$ stably seems difficult) \cite{2508.17231}, 
and scalar-tensor theories, i.e.\ modified gravity. 

On the other hand, there is a mythological creature called a 
chimera, where it is mostly a lion, but breathes fire, has 
a snake for a tail, and has a goat's head growing from its 
back. Such disparate elements make further elucidation (or 
prediction) difficult. If dark energy crosses $w=-1$ because 
of such a multiple field structure we have diminished hope of 
learning much physics. 

Models that may fall into either class includes theories 
with multiple unrelated fields, in which case there is often 
little one can say generally and little insight to gain, 
or where the dark energy is 
coupled to a matter sector, say. 
We will not pursue such interacting theories further since 
they tend often to have difficulty fitting data from large scale 
structure \cite{2506.02122}, but one can see 
\cite{2503.16415,2508.19101,2509.12335} for a few recent examples. 
In the next section we give a simple example where there is 
no interaction, but rather a misunderstanding of the matter 
sector. 

Alternately, one can cross $w=-1$ with a single field for certain 
theories beyond general relativity. For example, within the 
Horndeski class of modified gravity the presence of the 
$G_3$, $G_4$, or $G_5$ terms gives the possibility to cross $w=-1$. 
In any case, one can still define an effective dark energy 
density and pressure that enters the cosmic expansion history 
(though the perturbative sector will differ). 

Thus for all these cases we can return to the continuity 
equation 
\be 
\dot\rho_{\rm de}=-3H(\rde+P_{\rm de})=-3H\rde(1+\wde)\ , 
\ee 
and note that it implies that $\wde<-1$ gives dark energy 
density increasing 
with time, while if $\wde$ crosses $-1$ then at that instant 
one has a maximum (minimum) in the dark energy density as 
$\wde$ goes from less than $-1$ to greater (greater than $-1$ 
to less). The dark energy density can often be reconstructed 
from data with less noise than the dark energy equation of 
state (which involves a derivative of the density), and so 
nonmonotonicity of the dark energy density can provide a 
robust indication of crossing $\wde=-1$.

\section{Dark Sector Confusion} \label{sec:dm} 

It has long been appreciated that dark matter interactions and quintessence can give the resemblance of phantom-like dark energy \cite{Huey:2004qv,Das:2005yj}; more recent iterations include Refs.~\cite{2503.16415,2508.19101}. Here, we show in a simpler construct (based on Ref.~\cite{Liu:2025bss}) without interactions and with only $w \ge -1$ components, that misaccounting for a component of the cosmic fluid can lead to the appearance of a phantom that crosses back to quintessence. 

To proceed, consider the post-radiation era in which the energy budget consists of dark matter, a cosmological constant, and some ``X"-matter. There is very little of the X-matter, and it has an equation of state $w_{X} = -1 + a/a_0$. For this equation of state, the energy density evolution is $\rho_{X} = \rho_{X0} e^{3(1- a/a_0)}$. The total energy density is
\begin{equation}
    \rho_{tot} = \rho_{M} + \rho_{X} + \rho_\Lambda \label{eqn:true}
\end{equation}
where $\rho_M = \rho_{M0}\left({a_0}/{a}\right)^3$, and $\rho_\Lambda$ is a constant. To be specific, we can imagine the following breakdown: $\rho_{X0}:\rho_{M0}:\rho_{\Lambda} :: 0.01:0.29:0.7$.  

However, we cosmologists are unaware of this X, and model the expansion history using dark matter and dark energy only. For us, the total energy density is
\begin{equation}
    \rho_{tot} = \rho_m + \rho_{de} \label{eqn:model}
\end{equation}
where $\rho_m = \rho_{m0}\left({a_0}/{a}\right)^3$ and $\rho_{de}$ is to be determined. The breakdown is $\rho_{m0}:\rho_{de0} :: 0.3:0.7$.

The effective dark energy density is
\begin{equation}
    \rho_{de} = \rho_M - \rho_m + \rho_{X} + \rho_\Lambda. \label{eqn:rhode}
\end{equation}
We can easily determine the equation of state history of the dark energy by evaluating
\begin{equation}
    w_{de} = \frac{p_{de}}{\rho_{de}} = -1 - \frac{1}{3} \frac{d\ln \rho_{de}}{d \ln a}\ .
\end{equation}
Some quick math reveals
\begin{equation}
    w_{de} = - \frac{\rho_\Lambda + (1-a/a_0)\rho_{X}}{\rho_M - \rho_m + \rho_{X} + \rho_\Lambda}\ . \label{eqn_wDE}
\end{equation}
Next we will show that at late times, as $z \to 0$, $w_{de} \to -1$. And at an earlier time, around $z \sim 1$, $w_{de} < -1$.

First of all, we should point out that our clever choice, $0.01:0.29:0.3 :: \rho_{X0} : \rho_{M0} : \rho_{m0}$,  means $\rho_{X} + \rho_M - \rho_m|_0 = 0$. Now we can see that at $z=0$ or $a=a_0$, the numerator is just $-\rho_\Lambda$. The denominator is $+\rho_\Lambda$ because the other terms cancel out. So $w_{de} = -1$ today.

Second, at a redshift $z\sim 1$, the X energy density is negligible relative to the cosmological constant, so the numerator is dominated by the cosmological constant. Meanwhile the denominator is dominated by the cosmological constant minus a small bite, $\epsilon=\rho_m - \rho_M$, taken by the difference in matter energy densities. The equation of state is
\begin{equation}
    w_{de} \approx - \frac{\rho_\Lambda}{\rho_\Lambda - \epsilon} < -1\,.
\end{equation}
Hence, we have manufactured phantom dark energy out of normal components. This works because we have misaccounted for a fluid component that has some negative pressure (but $w>-1$) at early times, and vanishing pressure today. 
That negative pressure is enough to push the dark energy pressure over the phantom limit -- 
even though there is no phantom.  
Effectively, merging a misestimation of the matter density into the dark energy density takes a component with $w>-1$ but decreases 
$\delta H_1^2+\delta H_2^2$ in Eq.~(\ref{eq:wsum}) so as to give 
$\weff<-1$. 

\begin{figure}[b]
\centering 
\includegraphics[width=\columnwidth]{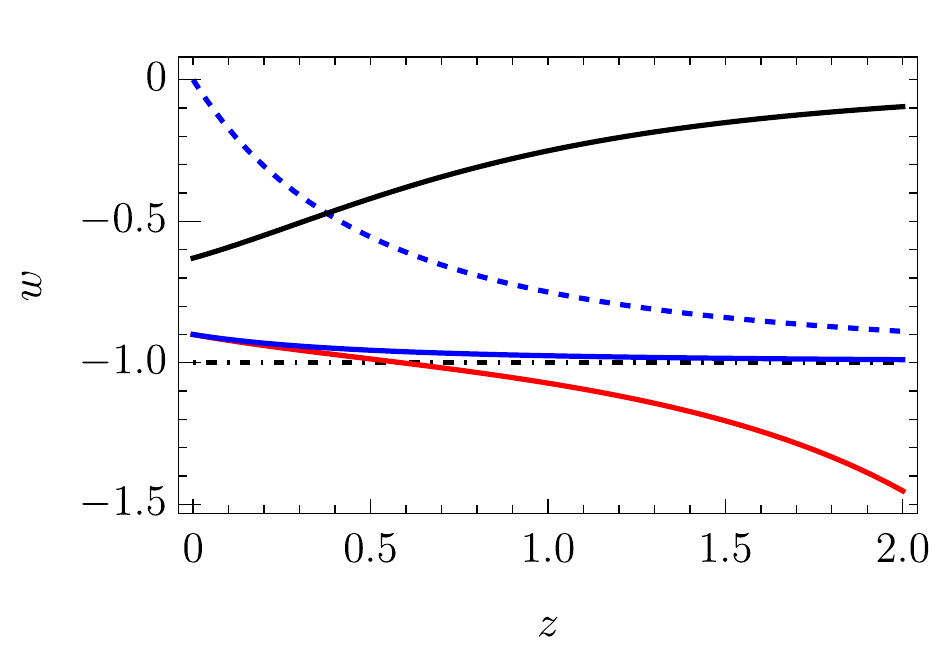} 
\caption{The equation of state for various components are shown: total (solid black), thawing quintessence (solid blue), thawing quintessence (X matter -- dashed blue), effective dark energy (solid red), cosmological constant (dot-dashed black). The phantom crossing occurs near $z \sim 0.5$.
} 
\label{fig:pm}
\end{figure}

We can arrange a phantom crossing if we replace $\rho_\Lambda$ by thawing quintessence, as a modest extension of Ref.~\cite{Liu:2025bss}. For the sake of quick modeling, we can use a thawing quintessence-like equation of state $w_Q =-1 + (1+w_{Q0})(a/a_0)^2$ where $w_{Q0}$ is the present-day value of the equation of state. If we set $w_{Q0} \sim -0.9$, for example, then we find the effective dark energy starts out phantom-like with $w_{de} < -1$ but crosses to the quintessence-like $w_{de} > -1$ near $z \sim 0.5$.

\begin{figure}[t]
\centering 
\includegraphics[width=\columnwidth]{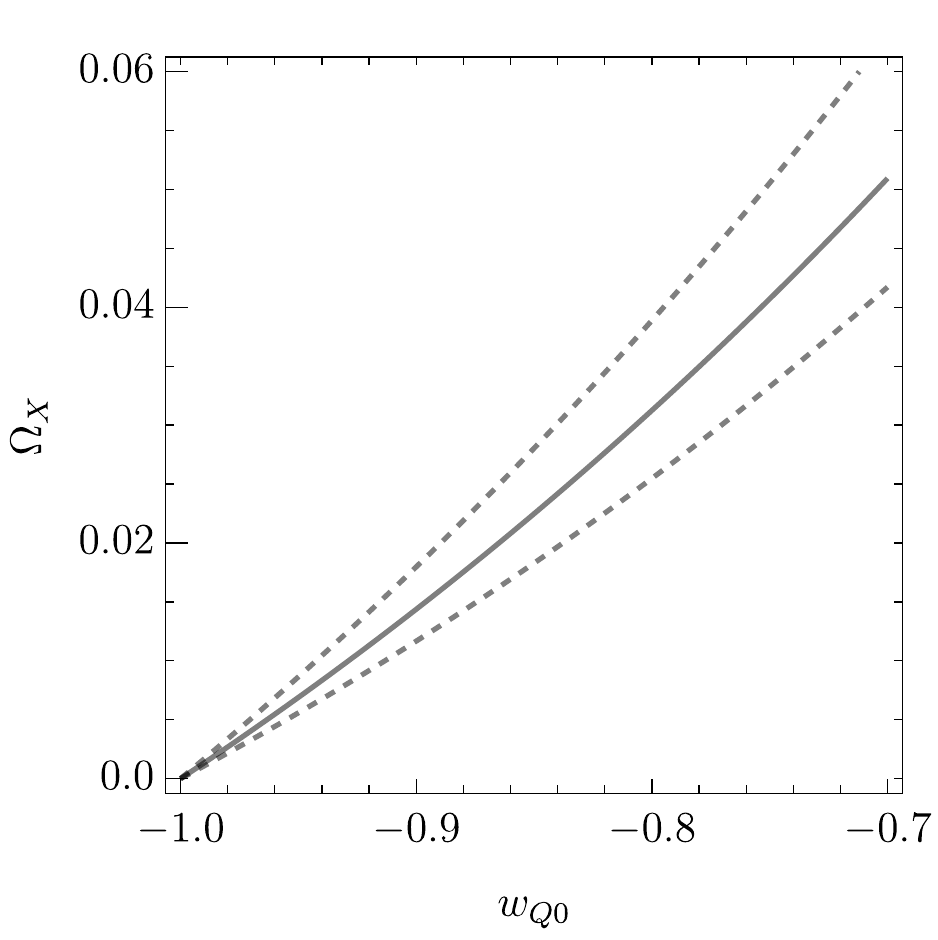} 
\caption{The range of parameters $\Omega_X$ and $w_{Q0}$ that yield a phantom crossing at $z = 0.5 \pm 0.05$ lies between the dashed lines. For simplicity, all other parameters are held fixed.
} 
\label{fig:Ww}
\end{figure}

Finally, we can replace the X-matter by a thawing quintessence field that has rolled part of the way down its potential to where $w = 0$. (Or we can consider the case where the field is fully thawed and its fast oscillations at the minimum of the potential contribute a pressureless fluid.) The results are essentially unaffected: the dark energy is phantom-like starting from $z \sim 3$; it makes a transition across the phantom line at $z \sim 0.5$, after which it is quintessence-like. The redshift of the crossing can be shifted by adjusting the parameters of this model. Since all components are essentially canonical scalar fields, the fluctuations are well behaved. The equation of state of each component is illustrated in Figure~\ref{fig:pm}.

The existence of a phantom crossing near $z=0.5$, as suggested by the current data, can be used already to constrain this toy model. For demonstration purposes, Figure~\ref{fig:Ww} shows the range of $w_{Q0}$ and $\Omega_X$ that yield a phantom crossing ($w_{de}=-1$) at $z=0.5 \pm 0.05$, all other parameters held fixed.

\section{Comparison to Current Data} \label{sec:data} 

\begin{figure}[t]
\centering 
\includegraphics[width=\columnwidth]{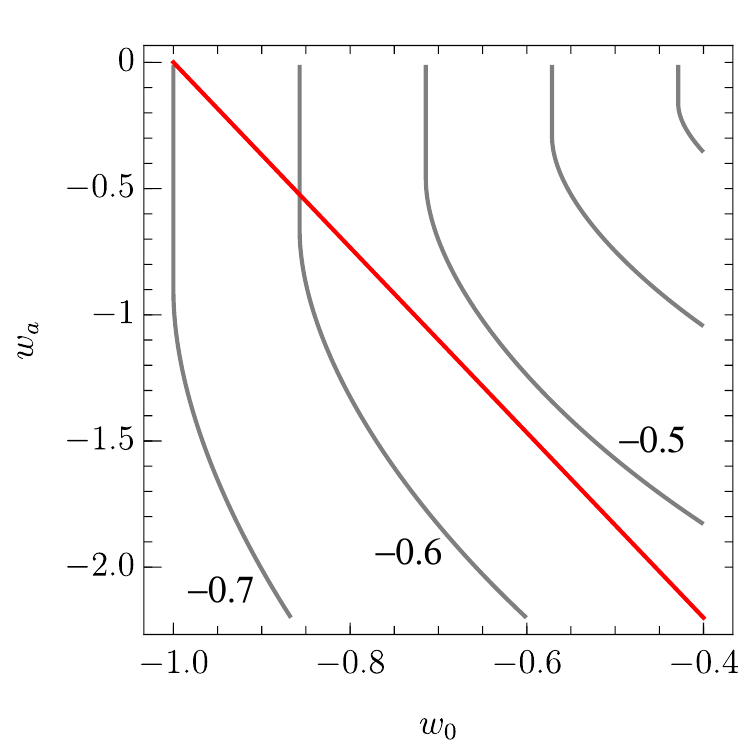} 
\caption{Our universe is far from violating NEC, 
with curves labeled by $\wtot$. The current constraints 
lie roughly along the red line indicating mirage dark energy, 
with the best fit to current data at $w_0\approx-0.7$, 
$w_a\approx1$, and hence $\wtot\approx-0.53$. Here 
$\om=0.3$. 
} 
\label{fig:necpast}
\end{figure}

Returning to the issue of NEC violation, 
and current data, we note that the global structure of spacetime is governed by the full stress-energy content of the Universe, and hence  $\rho_{\rm tot}+P_{\rm tot}$ is the relevant quantity. Equivalently, in equation of state terms, if we take matter (with pressureless equation of state) and put everything else into an effective dark energy, then the key question is whether 
\be 
\wtot=\wde\,\ode < -1\,? 
\ee 
Figure~\ref{fig:necpast} shows contours of $\wtot$ within the 
$w_0$--$w_a$ plane of dark energy parameters, where constraints from data are often plotted. 
(Recall that there $\wde(z)=w_0+w_a z/(1+z)$ for the dark energy 
equation of state at redshift $z$.) 
Current data constraints tend to lie along the ``mirage'' line 
given by the red diagonal line, with the current best fit about 
$w_0\approx -0.7$, $w_a\approx -1$, and the present matter density fraction $\om\approx0.3$. Such best fit values give 
$\wtot>-0.53$ for all times in the past, i.e.\ where observations 
can probe. No point on the mirage line has $\wtot<-0.7$. Thus  observations in no way point to NEC violation.

\begin{figure}[!th]
\centering 
\includegraphics[width=\columnwidth]{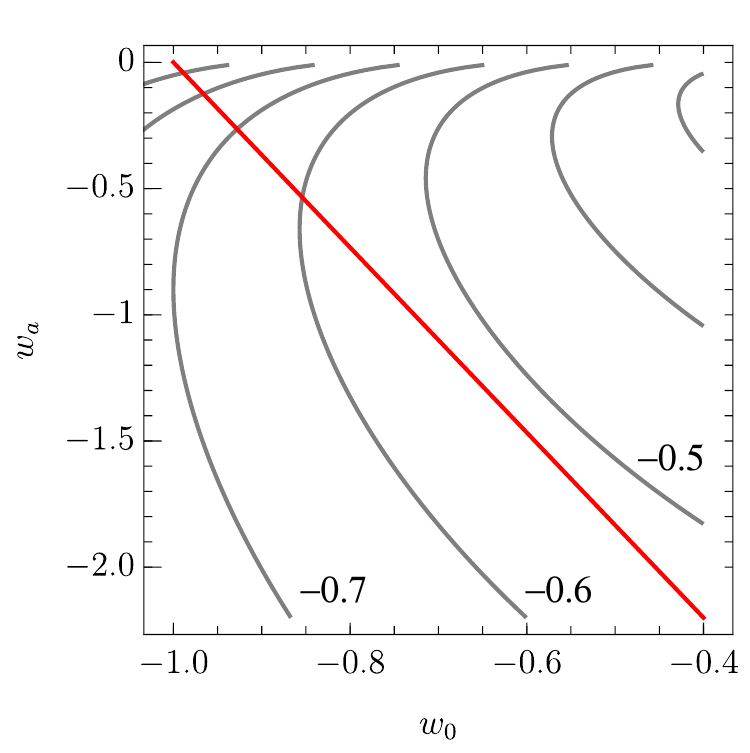} 
\caption{As Figure~\ref{fig:necpast}, but covering 
all times, including the future. Again, NEC will not be 
violated globally at any time. 
}
\label{fig:necall}
\end{figure}

We could ask whether NEC could be violated in the future 
(where we have no data). Of course we have no way 
of knowing what the behavior of dark energy in the future 
will be; if we naively adopt the $w_0$--$w_a$ into the future 
(for which it was never intended, since it is designed to 
match observables, not the equation of state per se) then 
we obtain Figure~\ref{fig:necall}, applicable to both past 
and future.

For the current best fit cosmology, the limit on 
$\wtot$ is basically the same as for Figure~\ref{fig:necpast}, 
as $w_{de}$ has already crossed $-1$ by the present. Even the 
upper endpoint of the mirage line (where the dark energy is 
the cosmological constant $\Lambda$) only eventually reaches $\wtot=-1$, 
but does not cross it. The 95\% confidence upper 
limit from current data gives $w_0\approx -0.80$, 
$w_a\approx -0.4$, lying along the mirage line, and only 
$\wtot>-0.65$. Thus there is no expectation of NEC violation 
ever in the cosmology indicated by current data.

\section{Conclusions} \label{sec:concl} 

The null energy condition has significant implications 
for the global structure of spacetime, but basically 
none for the cosmological model indicated by current data. 
At no time in the past (or possibly future) does the 
current cosmology violate even $\wtot\gtrsim-0.7$. While the 
dark energy density had a period where it increased with 
time (hence $\wde<-1$), this happens for 
various particle species, even on Earth, all the 
time and has no global effect. 

Despite NEC not being relevant to current cosmology, the 
presence of $\wde$ crossing $-1$ is very interesting and 
rules out many physical origins for cosmic acceleration. 
We broadly define viable theories into ``elephant'' and 
``chimera'' classes, i.e.\ where additional degrees of 
freedom live within a unified explanation or are 
disparate elements. Any theory must be compatible with the 
full array of cosmological data, including not only the 
cosmic expansion history but the growth of large scale 
structure and CMB perturbations. 

Measurements of the matter density at 
various times, e.g.\ through the CMB or 
tomographic large scale structure surveys, can further constrain ``confusion'' models such 
as presented in Section~\ref{sec:dm}. 
Further data elucidating both the high redshift ($z\gtrsim2$) 
and low redshift ($z\lesssim0.2$) behavior can provide 
valuable clues to dark energy properties. 
Accurate 
measurements of growth of structure, gravitational lensing, 
and gravitational wave sirens all enable different 
techniques for 
identifying the species of dark energy.

\acknowledgments 

We thank KASI for superb hospitality during the ``End of $\Lambda$?'' 
workshop, and participants Lisa Goh, Adria Gomez, Ryan Keeley, Shinji Mukohyama, 
Alessandra Silvestri, Shinji Tsujikawa, Masahide Yamaguchi for talks and conversations 
that helped shape this work.


\end{document}